\documentclass[12pt]{article}
\newcommand {\ve} [1] {\mbox{\boldmath $#1$}}
\begin{document}
\title{Comparative variational studies of $0^+$ states 
in three-$\alpha$ models}
\author{E.M. Tursunov$^{1,2}$, D. Baye$^1$ and P. Descouvemont$^1$ 
\thanks{Directeur de Recherches FNRS}
\\
$^1$ {\it Physique Nucl\'eaire Th\'eorique et Physique Math\'ematique, C.P. 229,} \\
{\it Universit\'e Libre de Bruxelles, B 1050 Brussels, Belgium} \\
$^2$ {\it Institute of Nuclear Physics, 702132, Ulugbek, Tashkent, Uzbekistan}
}
\date{}
\maketitle
\begin{abstract}
Three variational approaches, the hyperspherical-harmonics, 
Gaussian-basis and Lagrange-mesh methods 
involving different coordinate systems, are compared 
in studies of $0^+$ bound-state energies in 3$\alpha$ models. 
Calculations are performed with different versions of the shallow 
Ali-Bodmer potential (with and without Coulomb) and with the deep 
Buck-Friedrich-Wheatley potential. 
All three methods yield very accurate energies. 
Their advantages and drawbacks are evaluated. 
The implications of the disagreement between the obtained results 
and the experimental $^{12}$C energies in $3\alpha$ models 
with Coulomb interaction are discussed. 
\end{abstract}
\section{Introduction}
The analysis of the structure of halo nuclei such as $^6$He and $^{11}$Li 
has renewed the interest for accurate calculations of three-body systems. 
Various dynamical approaches to the three-body problem are known: 
the hyperspherical-harmonics method \cite{DTV98,TDE00}, the Faddeev 
coordinate-space method \cite{FJ-96,CFJ97}, the variational Gaussian-expansion method 
\cite{Ka-77,KKF89,TRK94,SK-98}, the Lagrange-mesh method \cite{HB-99,He-02}, \dots\  
Comparing the respective merits of such approaches is not easy because 
different groups usually make different choices for the conditions 
of calculation and even for the physical constants. 
Moreover the accuracy of the presented results is not always discussed. 

Therefore, we think that it is timely to make a comparison of three of the 
available methods which are in our expertise field. 
Our aim is to provide results with established high accuracy 
to which other methods can be compared, under exactly the same 
conditions. 
To this end, we have selected a simple example: 
the $3\alpha$ system with realistic local $\alpha$-$\alpha$ potentials. 
This example contains all the difficulties of this 
type of problem: (i) Coulomb interaction between all particles, 
(ii) orbital-momentum dependence of the nuclear interaction, 
(iii) occurrence of forbidden states in the two-body interactions, 
(iv) full symmetrization of the three-boson system. 
Methods of calculation can not always accurately treat all these 
difficulties. 
On physical grounds, this system is also interesting 
for several reasons. 

The three-$\alpha$-boson system has a long history in nuclear physics 
(see Refs.~\cite{No-70,TF-77} for discussions of early results). 
Accurate calculations could be performed with purely local 
$\alpha$-$\alpha$ realistic forces of the shallow type \cite{AB-66}. 
They clearly indicated that this model does not provide a realistic 
description of the $^{12}$C nucleus \cite{VW-71,MMS74}. 
Non-local two-body forces and/or three-body forces arising 
from the Pauli antisymmetrization play a non-negligible role. 
Strangely, a fair qualitative description of the bound spectrum 
is obtained when the Coulomb forces are ignored \cite{VW-71,MMS74}. 
Therefore three-body attractive forces can solve the problem, 
at least phenomenologically \cite{TF-77,FJ-96}. 

Deep potentials involving forbidden states are supposed  
to simulate the effect of the Pauli principle \cite{NKK71,BFW77}.
Performing accurate calculations with such forces turned out to be 
much more complicated because the two-body forbidden states 
must be eliminated from the three-body calculation. 
Otherwise, many unphysical states show up \cite{AFS92}. 
Attempts to solve that problem have led to contradictory results 
\cite{Ma-82,WN-87,AFS92,Tu-01}. 
Because of computer time limitations, whether the results 
obtained in Ref.~\cite{Tu-01} by one of us have reached convergence 
could not be established. 
It is thus timely to reexamine this question with several 
accurate methods and to establish firmly the $3\alpha$ bound states 
energies. 

The $3\alpha$ system has also recently gained a lot of interest 
as a possible example of a new phenomenon: 
the possible Bose-Einstein condensation of $\alpha$ bosons 
in the light nuclei $^{12}$C, $^{16}$O, \dots\ 
In Ref.~\cite{FHI80}, on the basis of microscopic calculations, it was proposed 
that the astrophysically significant $0^+_2$ state of the $^{12}$C nucleus 
at the excitation energy 7.65 MeV has a dominant relative $s$-wave structure 
as expected for a boson condensate.  
Moreover, a Generator Coordinate Method study \cite{THS01} of the $^{12}$C 
and $^{16}$O nuclei structure predicts the existence of near-threshold 
$n\alpha$ states which would be the analog for finite systems of an $\alpha$ 
condensation in infinite matter. 
However, recent variational calculations based on point-like bosons 
in $3\alpha$ and $4\alpha$ models with a local $\alpha$-$\alpha$ force 
obtain rather small $s$-wave components and lead 
to the conclusion that an interpretation as a Bose-Einstein condensation 
is too far from reality for these nuclei \cite{ST-02}. 
In this physical issue, it is important to work 
with as exact as possible wave functions. 
The accuracy of the results must be ensured to validate 
any physical interpretation. 

The aim of the present paper is to compare the results and merits of 
three different approaches of the $3\alpha$ bound-state problem 
for two types of effective $\alpha$-$\alpha$ interactions. 
The compared methods are 
the three-dimensional Lagrange-mesh method (LMM) \cite{HB-99,He-02}, 
the hyperspherical-harmonics method (HHM) on a Lagrange mesh \cite{DDB03}, 
and the variational method on a Gaussian basis (VGM) \cite{TRK94,Tu-01}. 
The LMM is under some conditions very accurate and can help 
calibrating the other two. 
The HHM is representative of variational techniques in hyperspherical coordinates. 
The VGM and its extension the stochastic variational method are versatile tools 
which can also treat larger numbers of particles. 

The selected effective $\alpha$-$\alpha$ interactions are 
the shallow AB potential of Ali and Bodmer with a strong repulsive
core \cite{AB-66} 
and the deep BFW potential of Buck, Friedrich and Wheatley \cite{BFW77}. 
For elastic $\alpha$-$\alpha$ scattering, these potentials are essentially 
equivalent \cite{Ba-87}. 
However, not only are their form factors very different 
but their use for a three-body system lead to quite different 
difficulties. 
Since the effects of the Pauli principle affecting the twelve nucleons 
can not be treated in an exact way in a $3\alpha$ model, 
one can try to simulate them with a microscopically founded potential. 
The deep BFW potential possesses unphysical bound states 
in the lowest $s$ and $d$ waves of relative motion to simulate the 
Pauli-forbidden states. 
However these forbidden states can not be kept in the three-body calculation 
because it becomes very difficult to find the physical states among 
the many eigenvalues of the Hamiltonian \cite{AFS92}. 
The forbidden states can be approximately eliminated from the solutions of 
the three-body Schr\"odinger equation with the method of orthogonalising 
pseudopotentials \cite{KP-78}. 
However, this $3\alpha$ model only takes into account two-body Pauli effects 
due to nucleon exchanges between two $\alpha$ particles. 
Using the full three-body Pauli projector is a very complicated problem 
\cite{Tu-01} which will not be considered here. 

The $3\alpha$ model is described in section 2, 
with emphasis on the potential choice. 
In section 3, the three methods of resolution of the three-body 
Schr\"odinger equation are summarized. 
In section 4, the results and the respective merits and limitations 
of the methods are compared. 
Some comments are made in section 5. 
Concluding remarks are presented in section 6. 
\section{Model and potentials}
The Hamiltonian of three identical bosons with mass $m_{\alpha}$ 
has the simple form 
\begin{equation} 
H = -\frac{\hbar^2}{2 m_{\alpha}} (\nabla_1^2 + \nabla_2^2 + \nabla_3^2 ) 
+ V(r_{23}) + V(r_{31}) + V(r_{12}),
\end{equation}   
where $\ve{r}_{ij} = \ve{r}_j - \ve{r}_i$ is the relative coordinate 
between particles $j$ and $i$. 
For deep two-body potentials with forbidden states in some partial waves, 
we use the method of orthogonalising pseudopotentials \cite{KP-78,Tu-01}. 
Each two-body potential $V(r_{ij})$ of the Hamiltonian is replaced by the 
corresponding pseudopotential 
\begin{equation}
\widetilde{V} (\ve{r}_{ij}) = V (r_{ij}) 
+ \Lambda \sum_{\rm f} \hat{\Gamma}_{ij}^{({\rm f})},
\end{equation} 
where the sum runs over the forbidden states. 
The constant $\Lambda$ must be taken large enough to push the forbidden 
states to high energies. 
The projector on a two-body forbidden state f is given by 
\begin{equation}
\hat{\Gamma}_{ij}^{({\rm f})} = \sum_{m_{\rm f}} 
| \varphi_{{\rm f} m_{\rm f}} \rangle 
\langle \varphi_{{\rm f} m_{\rm f}} |,
\label{0}
\end{equation}
where $\varphi_{{\rm f} m_{\rm f}} (\ve{r}_{ij})$ is the wave function 
of a forbidden state between $\alpha$ particles $i$ and $j$. 
As shown below, this wave function must be accurate. 
Because of the symmetry properties of identical bosons, 
only even partial waves of the $\alpha$-$\alpha$ potentials contribute. 

We use $\hbar^2/m_{\alpha} = 10.4465$ MeV fm$^2$ and $e^2=1.44$ MeV fm in all calculations. 
The Coulomb interaction is taken as 
\begin{equation}
V_{\rm C}(r) = 4e^2 {\rm erf\,} (\beta r)/r.
\end{equation}
The first model of the $\alpha$-$\alpha$ interaction used in this work is the 
AB potential d of Ref.~\cite{AB-66}, 
\begin{equation}
V_{\rm AB} (r) = V_1 \exp(-\eta_1 r^2) + V_2 \exp(-\eta_2 r^2) + V_{\rm C}(r).
\end{equation}
The $s$-wave parameters are $V_1 = 500$ MeV, $\eta_1 = 0.49$ fm$^{-2}$ and 
$V_2 = -130$ MeV, $\eta_2 = 0.225625$ fm$^{-2}$ for potential d. 
In the simplified $s$-wave version (ABd$_0$), these parameters are used 
in all partial waves 
and we choose $\beta = 0.75$ fm$^{-1}$ like in the BFW potential below. 
In the $l$-dependent potential (ABd), the parameters are as in the $s$ wave, 
except $V_1 = 320$ MeV for $l=2$, $V_1 = 0$ for $l \ge 4$ 
and $\beta = \sqrt{3}/(2 \times 1.44)$ in the Coulomb potential for all $l$ 
\cite{AB-66}. 
The second model makes use of the BFW potential of Ref.~\cite{BFW77}, 
\begin{equation}
V_{\rm BFW} (r) = V_0 \exp(-\eta_0 r^2) + V_{\rm C}(r), 
\end{equation} 
with $V_0 = -122.6225$ MeV, $\eta_0 = 0.22$ fm$^{-2}$ 
and $\beta = 0.75$ fm$^{-1}$.  
Both potentials describe fairly well the experimental phase shifts of the 
$\alpha$-$\alpha$ scattering for $l=0$, 2, 4 up to about 20 MeV. 
The BFW potential has three nonphysical bound states forbidden by the Pauli 
principle at the energies $E(0_1^+)=-72.625691755$ MeV, $E(0_2^+)=-25.618638588$ MeV 
and $E(2^+)=-22.000501732$ MeV. 
The $0_1^+$ state corresponds to the forbidden shell configuration $s^8$, 
while the $0_2^+$ and $2^+$ states correspond to $s^6p^2$. 
\section{Methods of resolution of the three-body Schr\"o\-dinger equation}
Basically, the three methods that we consider are all variational calculations 
of the three-body bound-state energies, but with different coordinate systems 
on one hand and different types of basis on the other hand. 
Two of the methods contain an additional Gauss-quadrature approximation. 
The three methods are based on a minimization of the expectation value 
of the Hamiltonian over linear or nonlinear variational parameters 
in a trial wave function. 
The comparison of their results is thus a significant test. 

The LMM \cite{HB-99,He-02} makes use of three-dimensional 
Lagrange basis functions, i.e. infinitely differentiable functions which 
vanish at all points of an associated mesh, with the exception of one. 
Because of the Gauss quadrature associated with the mesh, 
the potential matrix in the Lagrange basis is diagonal and its diagonal elements 
are the potential values at mesh points. 
Remarkably the Lagrange-mesh method appears to be as accurate as the corresponding 
variational calculation \cite{BHV02,DDB03}. 
However, it does not apply to $l$-dependent potentials. 

The Lagrange basis functions are defined in the system of perimetric coordinates 
\begin{equation} 
\begin{array}{l}
x = r_{12}-r_{32}+r_{13}, \\
y = r_{12}+r_{32}-r_{13}, \\
z = -r_{12}+r_{32}+r_{13},
\end{array}
\label{1}
\end{equation}
where $r_{ij} = |\ve{r}_i - \ve{r}_j|$ is the distance between two particles. 
The basis functions read 
\begin{equation} 
\Psi^{0^+}(x,y,z)= \sum_{ijk} C_{ijk} {\cal S} f_i(x/h) f_j(y/h) f_k(z/h)
\label{2}
\end{equation}
where ${\cal S}$ is the symmetrization projector and $h$ is a scaling parameter. 
The one-dimensional Lagrange functions read 
\begin{equation}
f_i (x/h) = (-1)^i (hu_i)^{1/2} \frac{L_N(x/h)}{x-hu_i} e^{-x/2h}
\label{3}
\end{equation}
where $L_N(u)$ is the Laguerre polynomial of degree $N$ 
and $u_i$ is one of its zeros, i.e. $L_N(u_i) = 0$. 
The interest of this approach is that the mesh equations resulting 
from the Gauss approximation are rather simple and only involve 
potential values at the $(hu_i,hu_j,hu_k)$ mesh points 
(see Ref.~\cite{HB-99} for details). 

In the following the calculations are performed 
with up to $N = 38$ mesh points for each of the three dimensions. 
The scaling factor $h$ is around 0.5 fm for $0_1^+$ and 1.2 fm for $0_2^+$ 
(the results are independent of the precise value of $h$). 
After symmetrization, the size of the largest matrix is $N(N+1)(N+2)/6 = 9880$ 
but computing times are short because the filling rate is about 10 \%. 

In the HHM \cite{DTV98,TDE00,DDB03}, the variational wave function is expanded 
over hyperspherical harmonics as
\begin{equation}
\Psi^{0^+}(\rho,\Omega_5) = \sum_{lKi} C_{lKi} {\cal S} {\cal Y}^{ll}_{K0}(\Omega_5) 
\hat{f}_i (\rho/h),
\label{4}
\end{equation}
where ${\cal Y}^{ll'}_{KL}$ are hyperspherical harmonics 
depending on five angular variables noted as $\Omega_5$. 
The hyperradial wave functions depend on the rotationally and permutationally 
invariant hyperradius $\rho$ and are expanded on regularized Lagrange-Laguerre 
basis functions 
\begin{equation}
\hat{f}_i (\rho/h) = (\rho/h u_i)^{3/2} f_i (\rho/h)
\label{5}
\end{equation}
where $f_i$ is given by Eq.~(\ref{3}).
Here also a Gauss approximation eliminates the need for calculations 
of potential matrix elements and significantly reduces the computing times 
(see Ref.~\cite{DDB03} for details and tests). 
One of the main advantages of this method is its validity for both bound-state 
and scattering problems. 

Here we use hyperspherical harmonics up to $K_{\rm max} = 30$ and $N = 30$ 
hyperradial mesh points with a scaling factor $h = 0.3$ fm. 
Because of symmetrization, the basis size is 
$N (K_{\rm max}+2) (K_{\rm max}+6)/16 = 2160$. 

The VGM \cite{Ka-77,KKF89,TRK94,Tu-01} is a high-accuracy method 
for the study of the structure of quantum-mechanical few-body systems. 
Its combination with stochastic methods \cite{SK-98} makes it applicable 
to few-body problems (up to six clusters at present \cite{VSL02}) 
in atomic, nuclear and quark physics. 
The variational wave function reads 
\begin{equation}
\Psi^{0^+}(\ve{x},\ve{y}) = \sum_{lij} C_{lij} {\cal S} 
[Y^l(\hat{\ve{x}}) \otimes Y^l(\hat{\ve{y}})]^{00} x^l y^l 
\exp(-\alpha_i x^2 -\beta_j y^2)
\label{6}
\end{equation}
where $\ve{x}$ and $\ve{y}$ are the Jacobi coordinates 
of any of the three possible sets, 
and the $\alpha_i$ and $\beta_j$ are non-linear parameters 
(see Ref.~\cite{Tu-01} for details and 
for the choice of non-linear parameters). 

Matrix elements of the Hamiltonian are calculated analytically 
for $l$-independent potentials with simple form factors 
such as ABd$_0$ (see \cite{TRK94,SK-98}). 
A numerical evaluation is necessary for the moments of potentials 
depending on $l$ or with a complicated form factor. 
However, in most cases, one can use recurrence formulas \cite{Tu-01} 
which allow to tabulate these moments. 
The high flexibility of a many-particle Gaussian basis makes it possible 
to describe several-particle configurations that are formed in the ground 
and excited states of multicluster systems \cite{TRK94}. 
For $l$-dependent potentials, extensions of the code are required 
with respect to Ref.~\cite{Tu-01}. 
If the potential term depends on a specific Jacobi coordinate $x$, 
it is necessary to perform a change of variable in the basis functions appearing 
in the matrix element and depending on another set of Jacobi coordinates 
in order to express them in the corresponding $(\ve{x},\ve{y})$ set. 
After an analytic integration over $\ve{y}$ and over the angular components 
of $\ve{x}$, the remaining integration over $x$ is performed numerically. 
A drawback of the VGM is the non-orthogonality of the basis 
which may restrict the basis size. 
In the VGM calculations below, the number of Gaussians is 680, 
except otherwise indicated. 
\section{Results}
The three numerical methods are used to determine energy values for the ground 
and lowest excited bound states  of the $3\alpha$ system with $J^{\pi} = 0^+$. 

The numerical results obtained with the ABd$_0$ potential are presented in Table 1. 
As in Ref.~\cite{ST-02}, let us first discuss calculations without the Coulomb term. 
The three methods yield close numbers which prove a high accuracy of all of them. 
The LMM is faster and more accurate. 
Its accuracy can be estimated by varying the scaling factor $h$ and 
the number $N$ of mesh points. 
>From these tests, all displayed digits should be correct. 
The LMM accuracy is much better for the ground state than for 
the excited state. 
The error of the VGM is about $2 \times 10^{-6}$ MeV.  
With 372 Gaussians, the results are $-5.12205$ and $-1.3523$ MeV 
while with 280 Gaussians they are $-5.1215$ and $-1.341$ MeV. 
The HHM accuracy is $10^{-3}$ MeV for the ground state but is 
less good for the excited state. 
To improve this accuracy, higher values of $K_{\rm max}$ should be used. 

With the Coulomb term in ABd$_0$, only one weakly bound ground state is obtained 
(see Table 1). 
The convergence of all methods is slower in that case. 
The accuracy is better than $10^{-8}$ for the LMM, $10^{-5}$ for the VGM 
and $10^{-3}$ for the HHM, in MeV. 
The $3\alpha$ energy is far from the experimental $^{12}$C energy $-7.275$ MeV 
\cite{AW-93} because off-shell effects are not well reproduced by the potential. 
Other Coulomb terms lead to similar results. 
With the Coulomb parameter $\beta$ from the AB paper, one finds $-0.61733886$. 
For the point Coulomb potential $-4e^2/r$, the energy is $-0.57238115$. 
With the value $\hbar^2/m_{\alpha} = 10.36675$ of Ref.~\cite{ST-02}, 
we obtain $-5.18093389$ and $-0.62106627$ MeV without and with point Coulomb, 
respectively, in excellent agreement with the values $-5.18$ and $-0.62$ 
obtained by these authors. 

Results with the full ABd potential are also presented in Table 1. 
The LMM does not apply to $l$-dependent potentials. 
We think that the accuracies of VGM and HHM are comparable to the ABd$_0$ case. 
They agree within $10^{-3}$ MeV. 
The binding energy is increased by about 1.3 MeV (without Coulomb) 
or 1 MeV (with Coulomb) with respect to the ground-state energy 
of the ABd$_0$ potential. 
In spite of this increase, the obtained energy remains not very realistic 
for $^{12}$C. 

Energies of the $3\alpha$ ground state calculated with the BFW potential 
including the Coulomb term are presented in Table 2 for the HHM and VGM. 
Because of the projection on forbidden states, the LMM can not be applied. 
To show the convergence of the orthogonalising-pseudopotentials method, 
we present numerical results for several values of the projection parameter 
$\Lambda$ ($10^5$, $10^6$, $10^7$, $10^8$ and $10^9$ MeV). 
It was shown in Ref.~\cite{Tu-01} that lower $\Lambda$ values 
can not fully eliminate the effect of the forbidden states. 
The VGM calculations are here performed with 280 Gaussians. 
Including more Gaussians leads to a numerical instability 
for large $\Lambda$ values. 
However, this rather small Gaussian basis already yields a good estimation 
for the ground-state energy as shown by the close agreement between the two 
variational approaches for each $\Lambda$ value. 
They agree within one or two percent. 
A fair convergence with respect to $\Lambda$ is obtained for values 
higher than $10^7$ MeV. 
Extrapolation with respect to $K_{\rm max}$ in the HHM yields $E = -0.30$ 
for $\Lambda = 10^7$, $-0.28$ for $10^8$ and $-0.27$ for $10^9$ (in MeV). 
The best result of Ref.~\cite{Tu-01} ($-0.283$ for $N = 7$ and $\Lambda = 10^8$) 
is not far from these values. 
The apparent lack of convergence observed in that work is due to a comparison 
with calculations with an insufficient accuracy on the forbidden states. 

The $3\alpha$ binding energy is 0.3 MeV smaller with the BFW potential 
than with the ABd$_0$ potential and 1.25 MeV smaller than with the ABd potential. 
The convergence of our results allows us to draw a conclusion that was 
not accessible with the basis sizes in Ref.~\cite{Tu-01}: 
a deep local potential accurately reproducing the $\alpha$-$\alpha$ phase shifts 
does not provide a better description of the $3\alpha$ system 
than shallow potentials. 
Independently of the choice of local potential, 
the $3\alpha$ model is not valid for the shell-model-like states of $^{12}$C. 
\section{A few comments on the $3\alpha$ model}
First we must emphasize the quality of the results obtained by Visschers 
and Van Wageningen more than 30 years ago \cite{VW-71}. 
For AB$d$ without Coulomb, they obtain $-6.37$ MeV for the ground-state 
energy, only 0.05 MeV about our result. 
This result is excellent taking into account the computer limitations 
of that time. 
With Coulomb, their result is less good as expected for the convergence 
of a smaller binding energy. 

Reliable results now exist for a deep potential. 
They confirm that the $3\alpha$ model with local forces is not able 
to reproduce the $^{12}$C ground-state energy. 
The deep-potential energy is even above the shallow-potential one. 
A similar effect is observed for the three-nucleon system \cite{KPF98} 
but not for the $^6$He halo nucleus \cite{DDB03}. 

An attractive three-body force can be used to cure this problem 
\cite{TF-77,FJ-96}. 
The effective three-body force employed in Ref.~\cite{FJ-96} 
depends however on the hyperradius $\rho$ and is specific 
for the HHM (see also Ref.~\cite{TDE00}). 
This type of form factor has little physical meaning. 

The origin of the weakness of the $3\alpha$ model 
should be understandable microscopically. 
Microscopic $3\alpha$-cluster models \cite{Ka-81,DB-87} provide results 
much closer to experiment. 
With an effective nucleon-nucleon interaction, they are able to reproduce 
fairly well the $\alpha$-$\alpha$ scattering and the $^{12}$C properties, 
simultaneously. 
In particular, the $^{12}$C ground state has a compact shell-model structure 
which can be reproduced by overlapping microscopic $\alpha$ clusters. 
On the contrary, the $3\alpha$-boson picture with local, deep or shallow, 
$\alpha$-$\alpha$ forces is far from reality. 
States with a compact shell-model structure are not satisfactorily 
described by the model. 
Moreover, the high sensitivity of the ground-state binding energy of the $3\alpha$ 
system to the description of the two-body forbidden states \cite{Tu-01} shows 
the importance of a correct treatment of the Pauli principle in 
cluster-model calculations. 

The success of 12-nucleon descriptions of the $3\alpha$ system 
has inspired a model which is apparently very similar to ours 
but which is based on a quite different philosophy. 
In this model local $\alpha$-$\alpha$ interactions are obtained by folding 
an effective nucleon-nucleon interaction and the forbidden states 
are those of the microscopic $\alpha$-$\alpha$ norm kernel. 
These forbidden states are simple oscillator states which are 
used in the pseudopotential (\ref{0}) in place of the exact bound 
states of the potential. 
This procedure leads to an overbinding of the $3\alpha$ system 
\cite{HKM97}, in opposition with our results. 

Using oscillator forbidden states with the BFW potential would be 
inconsistent. 
In the simple $3\alpha$ model that we have discussed, no information 
about the underlying structure of the $\alpha$ particle is available. 
The choice of the oscillator parameter in the forbidden states derived 
from a microscopic 12-body calculation would be quite arbitrary. 
We have tested this variant by replacing in the pseudopotential 
the exact forbidden wave functions of the BFW potential by 
oscillator wave functions. 
According to the choice of oscillator parameter, 
various ground-state energies can be obtained. 
For each choice, the energy is much lower than the energy $-0.26$ MeV 
given in Table 2 since the bound states of the BFW potential 
are only partly eliminated. 
In principle, with arbitrary forbidden-state wave functions, the 
pseudopotential technique can provide any result between the energy 
$-240.65$ MeV derived without elimination of forbidden states 
and the energy $-0.26$ MeV corresponding to full elimination. 
The more-or-less arbitrary choice of the wave functions of the forbidden 
state to be eliminated becomes a parameter of the model with which 
the experimental $^{12}$C binding energy could be reproduced or which 
can lead to overbinding as well as underbinding. 

The preceding discussion concerns local $\alpha$-$\alpha$ potentials. 
The role of non-locality in the interaction is not yet clearly evaluated. 
An attempt to elucidate this problem within the $3\alpha$ model has been performed 
in Ref.~\cite{FSM02} where a non-local interaction derived from a resonating-group 
microscopic calculation fitting the $\alpha$-$\alpha$ scattering provides 
a better-bound $^{12}$C ground state. 
However, the calculation and the elimination of forbidden states are directly 
performed with the $\alpha$-$\alpha$ $T$-matrix and the corresponding potential 
obtained after this elimination (which is presumably also non-local) 
is not available. 
\section{Conclusion}
To summarize, the energies of the $0^+$ ground and lowest excited states 
in the 3$\alpha$ model were evaluated in the framework 
of three variational approaches based on different coordinate systems. 
The LMM is faster and more accurate but is restricted to potentials 
which do not depend on the orbital momentum. 
The VGM is flexible and accurate but is more difficult 
to apply with $l$-dependent potentials. 
The non-orthogonality of the basis may cause problems. 
The slowness of convergence with respect to $K$ makes the HHM 
slightly less accurate than the other two. 
However, it can easily be adapted to $l$-dependent potentials 
and be extended to positive energies. 
The elimination of forbidden states with pseudopotentials shows 
convergence with respect to the parameter $\Lambda$ in contradiction 
with the conclusion of Ref.~\cite{Tu-01}. 
In order to reach convergence, the wave functions of the forbidden states 
must be very accurate. 

On the physical side, when used in a realistic way with the Coulomb interaction, 
both AB and BFW $\alpha$-$\alpha$ potentials yield poor results 
for the $3\alpha$ system. 
The 3$\alpha$ binding energies are too small compared with the $^{12}$C 
experimental value. 
The $^{12}$C ground-state structure can be described very well with 
{\it microscopic} $3\alpha$ models \cite{Ka-81,DB-87} but 
it can not be described with three-$\alpha$-boson models using realistic 
{\it local} interactions. 
The role of nonlocality in the two-body interaction still needs to be understood. 
It has recently started to be explored in Ref.~\cite{FSM02}. 
\section*{Acknowledgments}
This text presents research results of the Belgian program P5/07 on 
interuniversity attraction poles initiated by the Belgian-state 
Federal Services for Scientific, Technical and Cultural Affairs (SSTC).
One of the authors (E.M.T.) is supported by the SSTC.  
%

%
%\newpage
\begin{table}[hb]
\caption {$0^+$ energies (in MeV) of the $3\alpha$ system calculated 
with the $s$-wave (ABd$_0$) or $l$-dependent (ABd) Ali-Bodmer potentials 
of d type.}
\begin{center}
\begin{tabular}{|cccc|}
\hline
  state   &  LMM            &  HHM       &  VGM          \\ 
\hline
\multicolumn{4}{|c|}{ABd$_0$ without Coulomb}    \\ 
  $0_1^+$ & $-5.122093595$  & $-5.1219  $ & $-5.1220913$ \\
  $0_2^+$ & $-1.3606     $  & $-1.20    $ & $-1.3566   $ \\  
\hline
\multicolumn{4}{|c|}{ABd$_0$ with Coulomb ($\beta = 0.75$)} \\ 
  $0_1^+$ & $-0.58427008 $  & $-0.5836  $ & $-0.584266 $ \\
\hline 
\multicolumn{4}{|c|}{ABd without Coulomb}    \\ 
  $0_1^+$ &                 & $-6.423   $ & $-6.42285  $ \\
  $0_2^+$ &                 & $-1.92    $ & $-1.934    $ \\  
\hline
\multicolumn{4}{|c|}{ABd with Coulomb}   \\ 
  $0_1^+$ &                 & $-1.523   $ & $-1.523    $ \\
\hline
\end{tabular}
\end{center}
\end{table}
\begin{table}[hb]
\caption {$3\alpha$ ground-state energy calculated with the BFW potential 
for different values of the projection parameter $\Lambda$ (in MeV).}
\begin{center}
\begin{tabular}{|ccc|}    
\hline
$\Lambda$ & HHM & VGM \\ 
\hline
$10^5$ & $-0.638 $ & $-0.644$  \\  
$10^6$ & $-0.416 $ & $-0.426$  \\
$10^7$ & $-0.284 $ & $-0.288$  \\
$10^8$ & $-0.261 $ & $-0.263$  \\
$10^9$ & $-0.259 $ & $-0.261$   \\
\hline
\end{tabular}
\end{center}
\end{table}

\end{document}